\documentclass[twocolumn]{aastex701}
\usepackage{url}

\begin{document}

\title{A 2821 Star Optical SETI Survey using ESO HARPS archival data}
\author[0009-0008-7048-916X]{Benjamin Fields} 
\affiliation{Institute for Educational Advancement, Pasadena CA}
\affiliation{Wheaton College, Norton MA USA}
\email{bnfields72@gmail.com}

\author[0000-0002-6755-2710]{Jason C. Goodman}
\affiliation{Wheaton College, Norton MA USA}
\email{goodman_jason@wheatoncollege.edu}

\correspondingauthor{Benjamin Fields}
\email{bnfields72@gmail.com}
\date{August 2025}

\begin{abstract}
We examined archived observations of 2,821 stars taken by the high-resolution ESO HARPS spectrograph to search for potential narrow-band laser emissions from extraterrestrial sources. From one observation of each star, our search algorithm identified a total of 285 spectral peaks with line widths slightly larger than the instrument's point-spread function.  After eliminating  false positives (including cosmic rays, instrumental artifacts, and terrestrial airglow lines, we identified 8 sources worthy of follow-up observations. We then analyzed all 1,835 additional observations of these follow-up targets, looking for recurring signals. We found 1 additional unexplained candidate in this followup search, but no candidate spikes which repeated at the same wavelength as one of the initial candidates at a later time.  Further analysis identified one candidate as a likely faint airglow line. The remaining seven candidates continued to defy all false positive categories, including interference by LiDAR satellites and adaptive optics lasers from neighboring observatories. However, observations of other stars on the same night showed identical spectral spikes (in the telescope's reference frame) for four of these seven candidates -- indicating an as-yet unknown terrestrial source. This leaves 3 final candidates which currently defy the prosaic explanations examined thus far, show no indication of a terrestrial origin and therefore warrant further investigation. Two of these three candidates originate from M-Type stars and one of them originates from an oscillating red giant, so follow-up work will need to disentangle natural astrophysical stellar processes from potential SETI sources.
\end{abstract}

\section{The Case for Optical SETI Searches}
A majority of large scale SETI searches historically have concentrated on surveying radio wavelengths \citep{Gray_2017,OZMAII,LAMPTON1992189,Price_2020,PeterMa,Wright_2018}. One major advantage of this is that given their long wavelengths, radio waves can pass largely unobstructed through planetary atmospheres and the interstellar medium \citep{COCCONI_MORRISON_1959}. However, optical lasers provide a compelling alternative to radio, with many distinct advantages over them \citep{caseforopticalseti}. Extraterrestrial lasers could be used for communications, for propulsion, power transmission, or planetary defense.  

As a communications tool, lasers can transmit data at substantially higher rates than radio \citep{Tellis_2017,Kingsley}. They are also less affected by natural and anthropogenic interference than radio, which contends with many sources of RFI \citep{RFIpaper}, as well as naturally ocurring masers \citep{mendez2024arecibowowiastrophysical}. Because laser light is monochromatic, a laser could outshine its parent star within its particular wavelength band, showing up in the spectra as an unnaturally narrow emission line \citep{TOWNES_SCHWARTZ_1961,STANTON201992}. This has motivated numerous archival and dedicated searches for signatures of short-duration pulsed as well as continuous lasers \citep{Tellis_2017,Lipman_2019,Howard_2004,Shelley_Wright,Zuckerman_2023,Reines_2002,Maire_2019,Marcy_2021}.

A further motivation to search for lasers is that they have multiple potential applications beyond communication. Light sail propulsion systems are another potential source of detectable laser energy. \cite{Guillochon_2015} proposed to search for microwave light sail beams of the type proposed for interstellar mission concepts such as Starshot \citep{STARSHOT}. However, one could search for visible-light laser propulsion systems as well. Extraterrestrial lasers might also be used to transport power from one source to another within a solar-system spanning civilization, or used for planetary defense \citep{Bible_Johansson_Hughes_Lubin_2013}. Thus, lasers powerful enough to detect from earth-based telescopes may occasionally be directed toward us, either as a deliberate attempt to communicate or incidentally. In this work, we aim to identify any past detections of such events by searching archival spectroscopic data.

\section{Input Data}
\subsection{HARPS: An Ideal Instrument for Optical SETI?}
This survey uses archival reduced spectra from HARPS (High Accuracy Radial Planet Searcher), a high resolution spectrograph primarily built for exoplanet detection through the radial velocity method \citep{HARPS_specs}. It has a wavelength range of 378 nm-691 nm, with a gap in coverage between 530-533 nm, and a resolving power of 115,000 \citep{HARPS_specs} (which corresponds to about $10^{-2}$ \AA). The high resolution and large wavelength range of HARPS makes it ideal for searching for narrow band laser emissions. 

\subsection{Target Selection}
The HARPS database includes observations not just of stars, but solar system objects as well as  astrophysical exotica such as quasars and supernovae \citep{trifonov_2020}.  While ETI may not be constrained to stellar systems (\citep{CIRKOVIC2018289}), in this survey we focus our search on stellar spectra. However, we do not limit our search to particular spectral types or systems with known planets or on the basis of potential habitability.

Our search list comes from  \cite{trifonov_2020}, who provided a database of stellar targets observed by HARPS at least three times, which was compiled for the purpose of radial velocity searches for exoplanets.  This database includes 212,000 observations of 2,821 stars observed by HARPS through 2019, as well as distances and spectral types for each target. Our search algorithm focuses on reduced spectra for the initial search, followed by manual analysis of raw CCD data to minimize the impact of false positives.

\begin{figure}
    \centering
    \includegraphics[width=\columnwidth]{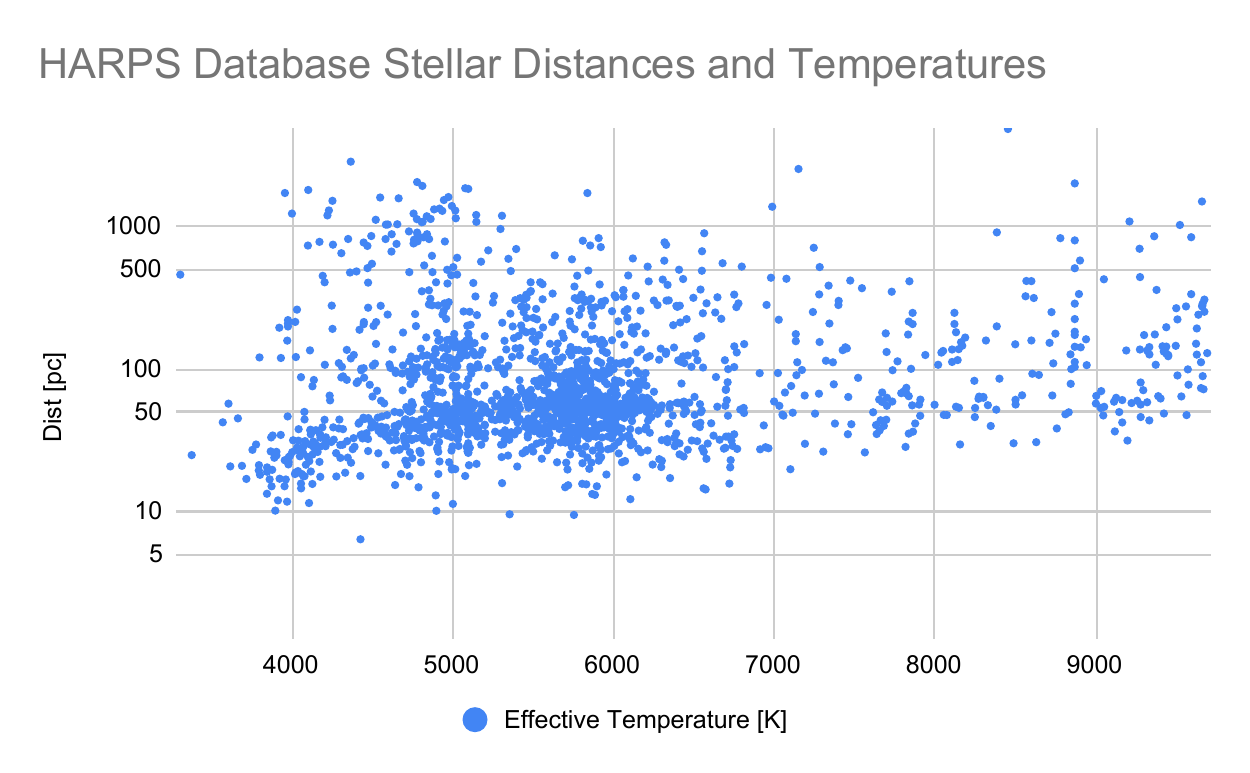}
    \caption{Temperatures (x-axis, Kelvin) and distances (y-axis, parsecs) of the stellar population targeted by this study. }
    \label{fig:starlist}
\end{figure}

As can be seen in Figure \ref{fig:starlist}, this search covers a wide range of stellar temperatures, with cooler stars of types F, G, K, and M with surface temperatures less than 7000 Kelvin are the most represented. The vast majority of stars surveyed  are within a distance of 500 parsecs or less \citep{trifonov_2020}. Selecting nearby stars is advantageous for both HARPS planet detection mission and our laser detection project, as it makes faint signals easier to detect. The spectral types confer a similar advantage, as dimmer stars provide less background light that the laser would need to outshine \citep{Howard_2004}.

In the initial search described in this work, we analyzed only one observation from each of these targets (generally, the first one returned by the ESO HARPS data archive search tool at \url{https://archive.eso.org/cms/data-portal.html}). For the handful of stars which yielded compelling candidates from the initial survey, we analyzed all subsequent observations. Future work will expand this search to encompass every available stellar spectra in the HARPS database.

\section{Search Methodology}
We subjected the targets to increasing levels of scrutiny, from automated search tools to detailed manual analysis.  We organize our search by tiers: candidates that pass the Tier 1 tests become Tier 1 candidates, which proceed to Tier 2 analysis, and so on.

The structure of our search is as follows:
\begin{itemize}
    \item Tier 1: Automated Search and Basic Airglow Detection (Section \ref{s:automated})
    \item Tier 2: Manual Assessment of Other False Positives (Section \ref{s:falsepositive})
    \item Tier 3: Curve-Fitting Analysis (Section \ref{s:curvefit})
    \item Interlude: Searching for Repeat Occurrences
    \item Tier 4: Airglow Reanalysis (Section \ref{s:airglow_reanalysis})
    \item Interlude: Detailed inspection of the HD127423 Candidate
    \item Tier 5: Doppler Detection of Unknown Terrestrial Sources (Section \ref{s:doppler})
\end{itemize}

\subsection{Tier  1: Automated Search Algorithm}
\label{s:automated}

Our search algorithm must differentiate between possible laser signals, natural emission lines from the star or Earth's atmosphere, and artifacts created inside the HARPS instrument.  We use the spectral width of the feature as one of our distinguishing parameters. The point-spread function (PSF) describes the spectral resolution of the instrument, and shows how the instrument's optics would spread out a monochromatic light source to a finite spectral width at the instrument's CCD cameras. The full width half-maximum (FWHM) describes the the observed width of a spectral feature.  Since we are searching for near-monochromatic signals emitted from space, we should look for spikes in the spectral data whose FWHM is similar to the instrument's PSF.

Observed spectral peaks with a FWHM narrower than the PSF cannot represent light that passed through the instrument's optics, and are likely to be cosmic ray hits or other instrumental artifacts. Conversely, broader spikes are likelier to be from natural sources due to thermal doppler broadening from relatively high temperature astrophysical processes.

Our automated search algorithm follows a similar approach to previous archival optical SETI surveys \citep{Tellis_2017,Lipman_2019,Marcy_2021}. First, it establishes a baseline ``continuum'' stellar flux describing the star's overall spectrum by performing a running median filter with a width of 100 pixels (1 angstrom).  This is wide enough to smooth away many natural absorption and emission lines, but narrow enough to minimize the data discarded at the extreme ends of the spectrum.  It also calculates a running standard deviation using the same window width.

Next, the algorithm searches for spectral data points more than 3.5 standard deviations above the running median.  This threshold is chosen to be as low as possible to maximize sensitivity, while minimizing the number of false positives resulting from random data noise.  

Our algorithm looks for multiple spectral data points in a row that exceed this threshold, within both a minimum and maximum spike width.  The HARPS instrument itself has an average PSF of approximately 4.1 pixels \citep{HARPS_specs} or 0.04 angstroms (though this varies with wavelength), so monochromatic signals from space should be spread out at least this much. This varies by wavelength, but serves as good general minimum value for our algorithm. Thus, we look for \emph{at least four pixels in a row} at or above the 3.5 sigma threshold.  If the pixel values were independent Gaussian random values, the probability of four in a row rising above the 3.5$\sigma$ confidence interval would be about $5000^{-4}=10^{-15}$, which means we would be unlikely to see a single false positive from random noise anywhere in the measurements encompassed in our dataset. Of course, the instrument noise is not Gaussian and far from independently random, but this demonstrates that the probability of a random ``4-in-a-row'' spike is very small, and in practice our algorithm did not find any signals whose properties were consistent with misinterpreted random noise.

The algorithm also ignores any spikes with more than 60 spectral data points in a row above the threshold.  This maximum criterion helps to weed out natural stellar emission lines, which tend to have wider peaks due to thermal doppler broadening, and also prevents spurious ``detections'' at the edge of the HARPS instrument's inherent data gap. The 100-point running median filter used to establish the continuum cannot filter out spectral features wider than about half the filter width, so our algorithm is insensitive to such wide spikes anyway.  Spikes wider than 4 spectral data points above the threshold but narrower than 60 are flagged as initial potential laser candidates and subjected to further review (See Section \ref{s:falsepositive}).

\begin{figure}
    \centering
\includegraphics[width=\columnwidth]{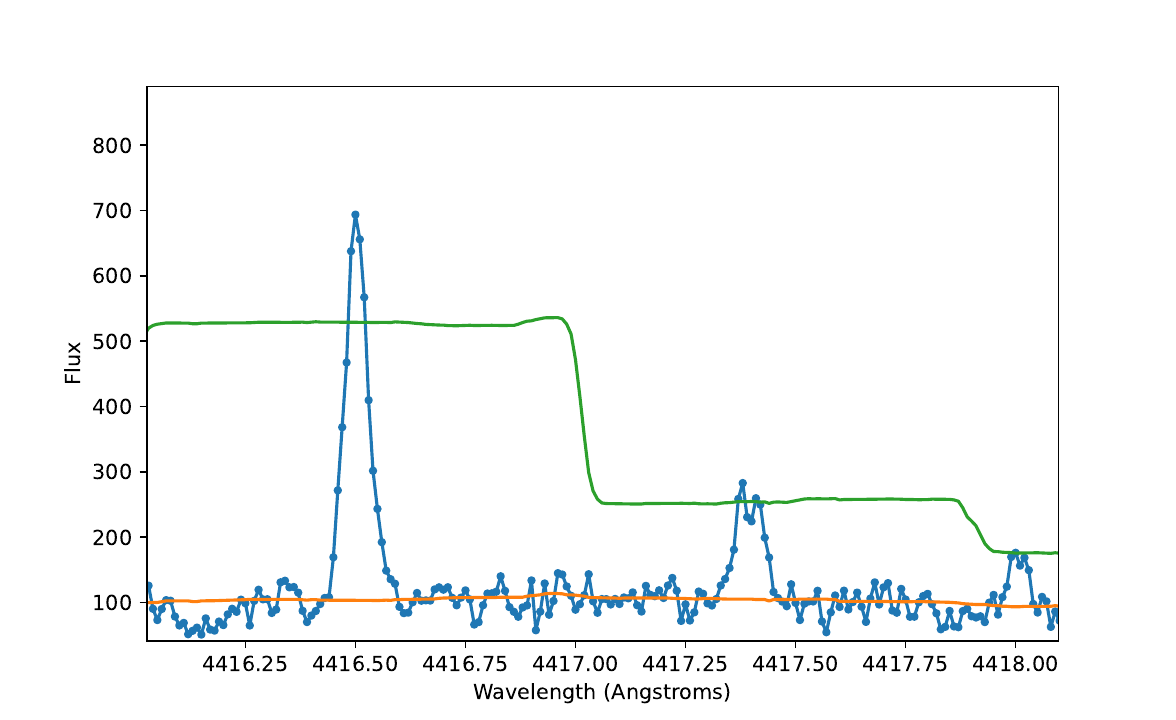}
    \caption{A representative  candidate signal from Proxima Centauri (later found to be part of a family of natural stellar emission lines). (HARPS OBSERVATION ID: \path{ADP.2014-09-25T15_36_33.623.fits}). Blue: spectral data, orange: 100-point running median, green: noise threshold (median + 3.5 x local running standard deviation).}
    \label{fig:algorithm-example}
\end{figure}

To visualize the algorithm, we present an example candidate signal from Proxima Centauri (Figure \ref{fig:algorithm-example}).  Note that the data spikes themselves affect the local standard deviation, and thus the noise threshold. The leftmost spike has more than 4 pixels in a row above the threshold, and so would be flagged as a candidate signal. The other two spikes have fewer than 4 pixels above the threshold, and so would not be flagged.

\subsubsection{Automated Airglow Detection}
\label{s:Airglow}
\begin{figure*}
    \centering
    \includegraphics[width=\textwidth]{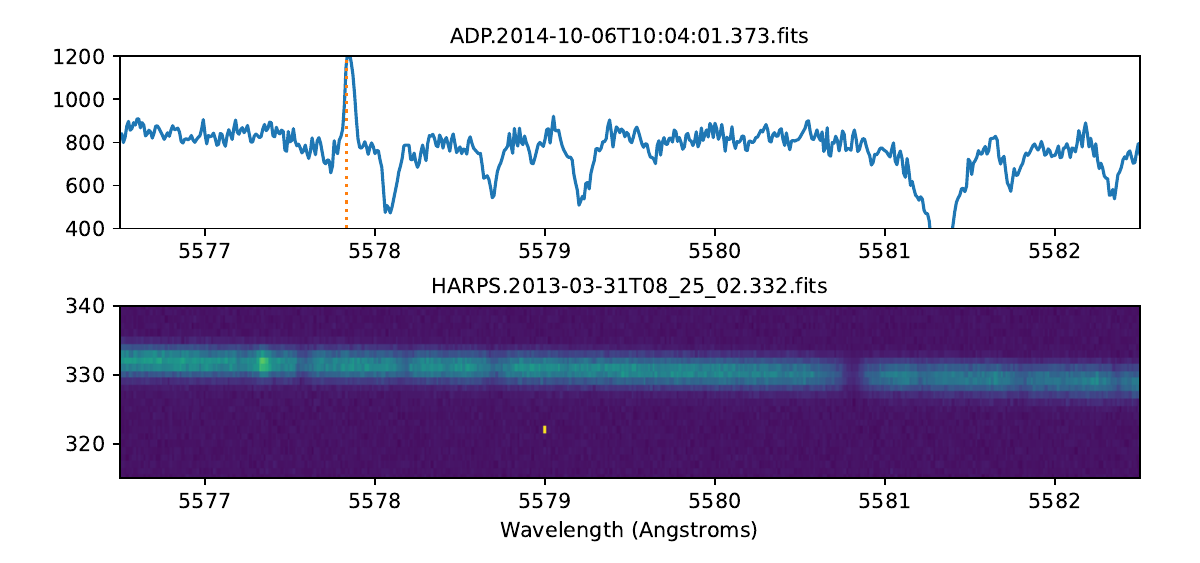}
    \caption{An example of a commonly recurring airglow line at 5577 \AA (Observation Date/Time: 2013-03-31, 08:25:02.332). Top panel: reduced spectrum (the red dashed line indicates the spike detected by the automated algorithm).  Bottom panel: subsection of the corresponding CCD image for this wavelength range. The stellar spectrum is the pale blue band. At the location of the detected spike, the CCD image shows a visible corresponding bright spot.  As in all figures of this type throughout this work, there is a small wavelength offset between the upper and lower panels because the HARPS pipeline applies a doppler shift to the reduced spectrum.}
    \label{fig:airglowexample}
\end{figure*}

Atomic transitions in Earth's atmosphere (airglow) are a major source of false positives in our data.  Since these emission lines come from heavy atoms in a cool low-pressure atmosphere, their spectral width is near the instrumental PSF. Therefore, they are difficult to distinguish from lasers using linewidth alone. However, night sky airglow emissions only occur at specific wavelengths. Our initial search algorithm ignores any spikes detected in a 1-angstrom-wide window surrounding several of the most prominent airglow line wavelengths \citep{Luger_2017}. These `prohibited wavelengths' are shown in Table \ref{table:AirglowWavelengths}. For a visual example of an airglow line, see Figure \ref{fig:airglowexample}.

\begin{table}
\begin{center}
\begin{tabular}{|c|c|} 
 \hline
 Element & Wavelength \\ 
 \hline
 He & 3889 \AA \\ 
 \hline
 H & 4861 \AA, 6563 \AA \\ 
 \hline
 N2 & 5198 \AA, 5200 \AA \\ 
 \hline
 OI & 5577 \AA \\
 \hline
 OI-III & 6300 \AA, 6364 \AA \\
 \hline
 Na & 5889 \AA , 5896 \AA \\
 \hline
 \end{tabular}
\end{center}
 \caption{Major airglow lines ignored by our automated search algorithm.}
 \label{table:AirglowWavelengths}
\end{table}

Unfortunately, this means that any lasers transmitted at those wavelengths would not be detected by our search tool.  Doppler shift information could be used to distinguish between terrestrial and stellar sources (see Section \ref{s:unexplained}), but the initial search simply rejected all spikes within 1 {\AA} of the brightest airglow lines.

Our Tier 1 automated search, based on the above criteria (4-60 pixels in a row above threshold, with several known bright airglow lines excluded) yielded 285 ``spike'' candidates from our 2,821 stars. Additional manual analysis was performed to exclude other false positives, as discussed in Section \ref{s:falsepositive}.  A later step (Tier 4) also included a reanalysis of airglow lines using the comprehensive airglow line database compiled by \cite{UVES_database}, for any rarer/fainter airglow lines which our automated filter did not account for: see Section \ref{s:airglow_reanalysis}. 

\subsection{Tier 2: Eliminating Other False Positives}
\label{s:falsepositive}
Once the automated search algorithm identified Tier 1 targets, our Tier 2 analysis applied a combination of automated and manual criteria to identify potential false positives and distinguish potential technosignatures from prosaic phenomena.  In addition to airglow lines (considered in Tier 1 (Section \ref{s:Airglow}) and reconsidered in Tier 4 (Section \ref{s:airglow_reanalysis})), we looked for three other categories of false positive: cosmic rays, stellar emission lines, and calibration lamp bleedthrough.

\subsubsection{Cosmic Rays}
\label{s:cosmic}
Cosmic rays can collide with the HARPS instrument's CCD sensors, scattering electrons and creating false "spikes". Since these spikes are not created by light passing through the instrument, many will be narrower than the instrument's PSF, and will be filtered out by the algorithm's ``4-in-a-row'' criterion.  However, in some cases high-energy particles can slice through multiple pixels at an angle, forming meteor-like tracks in the raw CCD image and creating a spectral peak more than 4 pixels wide.  To detect such cases, we manually inspected the raw CCD images for each automatically-detected peak to visually identify patterns characteristic of cosmic rays, utilizing a methodology similar to \cite{Tellis_2017}. 

We identified likely cosmic rays by visual inspection using the following criteria:
\begin{enumerate}
    \item a long, narrow track on the CCD image that was not aligned with the spectrometer's entry slit, or
    \item a CCD signal that spills over into parts of the image that are not illuminated by the starlight passing through the spectrometer's entry slit, or
    \item a signal that does not fully occupy the width of the region illuminated by the entry slit.
\end{enumerate}
We manually classified a detection as a ``likely cosmic ray'' if it had any of these features.

\begin{figure*}
    \centering
    \includegraphics[width=\textwidth]{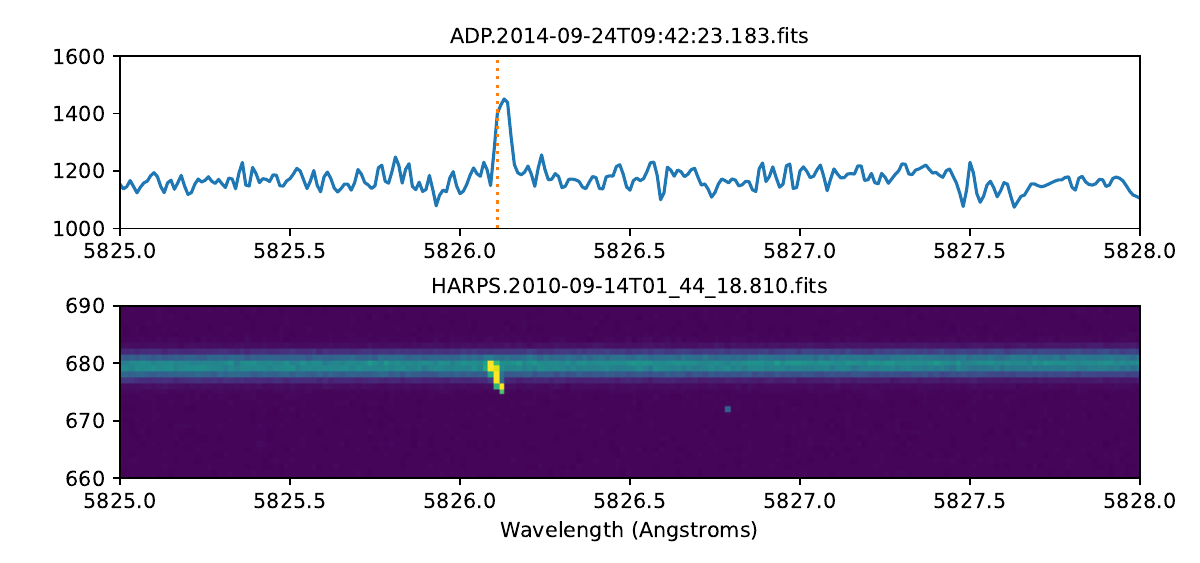}
    \caption{An example of a cosmic ray (Observation Date/Time: 2010-09-14, 01:44:18.810). Top panel: reduced spectrum (the red dashed line indicates the spike detected by the automated algorithm).  Bottom panel: subsection of the corresponding CCD image for this wavelength range. The stellar spectrum is the pale blue band. At the location of the detected spike, the CCD image shows a long track which spills over outside the band of starlight and into surrounding CCD pixels.  }  
    \label{fig:Cosmic_ray_example}
\end{figure*}

\subsubsection{Calibration Lamp Bleed-through}
\label{s:bleedthrough}
\begin{figure*}
    \centering
    \includegraphics[width=\textwidth]{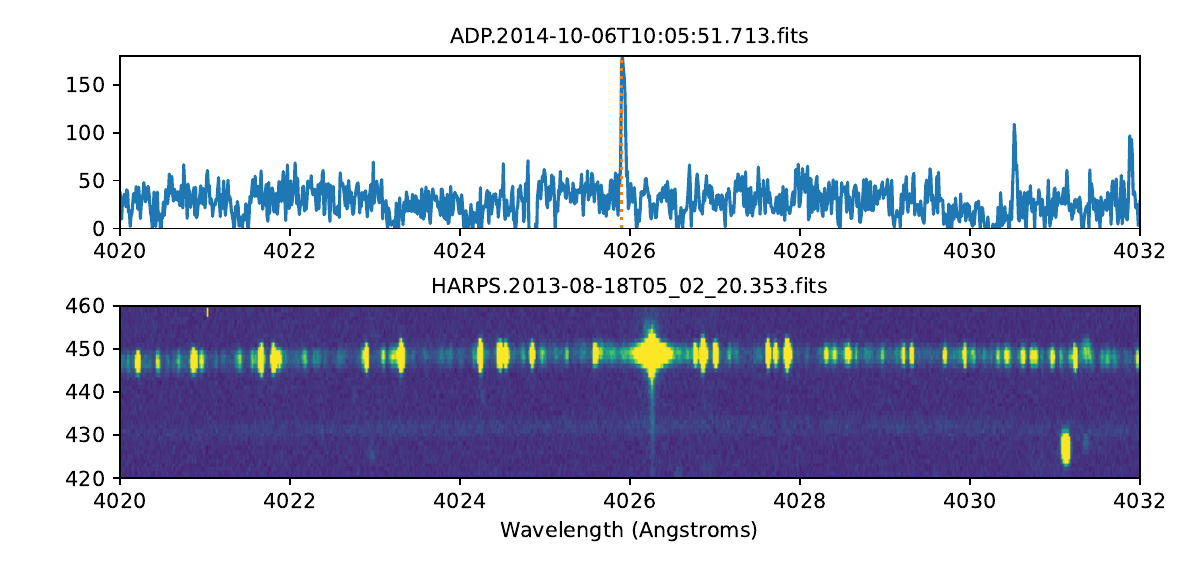}
    \caption{An example of calibration lamp bleed-through (Observation Date/Time: 2013-08-18, 05:02:20.353). Top panel: reduced spectrum, bottom panel: raw CCD image. The upper band of bright spots in the CCD image is from the calibration lamp; the lower band is starlight.  Charge from an oversaturated emission line in the calibration lamp has bled across to contaminate the stellar spectrum.  As in all figures of this type throughout this work, there is a small wavelength offset between the upper and lower panels because the HARPS pipeline applies a doppler shift to the reduced spectrum.}
    \label{fig:bleedthrough}
\end{figure*}

 The HARPS instrument contains a calibration lamp, an artificial reference spectrum imaged in the same CCD frame as the stellar spectrum \citep{HARPS_specs}. In some observations, the calibration lamp was left on during the observation, either deliberately or by mistake.  This led to another class of false positives, ``calibration lamp bleed-through'' (see Figure \ref{fig:bleedthrough}).
 
In the example shown in the figure, the calibration lamp's bright spectral peaks (seen in the upper part of the CCD image) ``bleed through'' vertically, creating a spurious spike in the stellar spectrum below.  Since the width of these spikes is close to the instrument's PSF, this category of false positive cannot be identified automatically in processed spectra and was identified by manual inspection of the CCD images.

\subsubsection{Natural Stellar Emission Lines}
\label{s:stellaremission}
Natural emission lines lines from stars could also be detected by our spike-searching algorithm.  These natural sources have two features that might distinguish them from lasers.  First, due to the doppler broadening effect, emission lines from hot low-molecular-weight gases tend to have spectral peaks much broader than the HARPS instrument's PSF. However, emission lines from cooler stars, or those from heavier elements, may be narrower \citep{Livadiotis_2018}.
 
Secondly, natural stellar emission lines rarely show up in isolation. Most stars do not have strong emission lines; the ``active'' stars that do tend to have many of them. Typically, a stellar emission line will occur as part of a “series” in which many different atomic transitions of one or more elements are excited in the atmosphere of an active star.  Thus, the number of spectral peaks is another criterion that can distinguish between natural stellar emission lines and lasers. We filtered out cases where more than 4 spikes were detected in a single observation, categorizing these as likely natural stellar emission lines.  Of course, an alien technological source might use multiple laser wavelengths--however, it remains a telltale characteristic of natural stellar activity \citep{Livadiotis_2018}, especially when coupled with relatively wide peaks due to the thermal doppler broadening that the heat of a star would likely produce. Regarding the doppler broadening, there can be exceptions. As noted in \citep{ProximaCentauri}, some M-type stars with active flaring activity can sometimes produce emission lines  with  narrow and almost 'laser-like' linewidths due to their relatively low temperatures and high metallicity. However, in such cases the flaring activity will produce a multitude of emission lines and any spikes will not appear in isolation--a scenario addressed by our  criteria.

\begin{figure}
    \centering
    \includegraphics[width=\columnwidth]{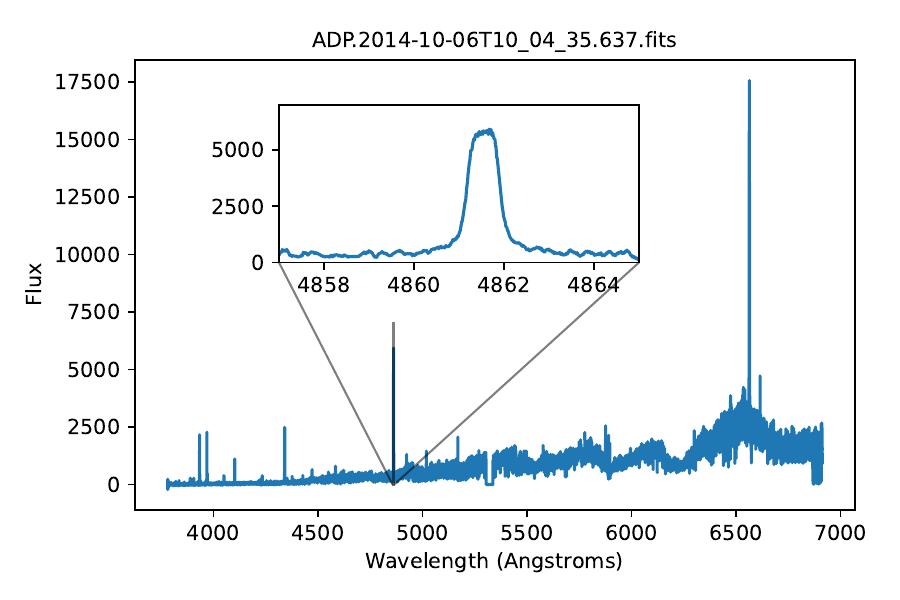}
    \caption{Reduced spectrum of an eruptive variable star, GJ234. Numerous spikes are visible, as is typical of natural stellar emission-line activity from eruptive stars like this one.  Inset: Zoom-in on one of approximately 10 spikes observed from 
    this star. These emission features are much broader than the HARPS instrument's point spread function, as is typical of natural stellar emission lines.}.
    \label{fig:eruptivevariablestar}
\end{figure}

For an example of a spectra which matches our criteria for natural stellar emission lines, see Figure \ref{fig:eruptivevariablestar}  from eruptive variable star GJ234.

Manual inspection of CCD frames let us categorize 141 of our candidates as cosmic rays and 10 as calibration lamp bleedthrough. Automated filters eliminated the most prominent/recurrent airglow lines {\em a priori}, and 89 candidates appeared in clusters indicative of stellar emission lines. This left us with 41 Tier 2 candidates still unexplained.

\subsection{Tier 3: Curve-fitting algorithm}
\label{s:curvefit}

Although our manual search was effective at spotting many cosmic rays, some will create trails that do not show diagonal streaks or extend outside the stellar spectrum, and will thus not be visually distinctive.  Other unidentified instrumental artifacts might also create false positives.  To rule these out, our Tier 3 analysis fit a Gaussian curve to each of our candidate spikes, and considered its linewidth.  Line features caused by light passing through the HARPS optics will always have a line width greater than the instrument's point-spread function (PSF), while cosmic rays or internal artifacts may not. Although we attempted to mitigate this by setting a minimum number of pixels above the threshold in our initial search algorithm (See \ref{s:automated}), the PSF varies by wavelength and more precise methods are needed to accurately ascertain linewidth.

HARPS has a resolving power \citep{HARPS_specs} of 115000, meaning that the instrument's PSF at a given wavelength is given by 

\begin{equation}
    \delta\lambda = \frac{\lambda}{115000}
    \label{eq:psf}
\end{equation}

The linewidth analysis was performed with \texttt{astropy.specutils} (\url{https://specutils.readthedocs.io/en/stable/}). After subtracting the background continuum, we fit a Gaussian curve to each of the 41 remaining candidate spikes and used it to estimate the spike's Full-Width Half-Maximum (FWHM).

If the best-fit width of a spike is greater than or equal to the instrument's PSF (Equation \ref{eq:psf}), the candidate is likelier to be a genuine SETI detection. Otherwise, the candidate signal possibly originates from within the instrument and is therefore likelier to be a cosmic ray or instrumental artifact.

To validate this criterion, we tested it on two types of false positives: one consisting of Tier 1 candidates which we documented as cosmic rays in our initial manual investigation, the other candidates that matched known airglow lines.  Since airglow lines originate from cool, heavy atoms, they have line widths narrower than the instrument's PSF, and therefore serve as good approximations of laser-like  signals.  If our curve fitting algorithm was working as expected, we should expect that a majority of cosmic rays would fall below the predicted linewidth, while a majority the airglow lines would exceed the predicted linewidth.

This is the exactly what test showed.  Of 206 spikes manually identified as cosmic rays, 141 (68\%) fell below the predicted FWHM.  Of 188 spikes matching the wavelengths of common bright airglow lines, 15 (7.9\%) fell below the predicted FWHM, with the rest exceeding it. The cosmic rays which passed the test are likely due to particles slicing through multiple pixels, thus appearing to be at or above the instrumental PSF. However, the 15 airglow lines which failed the test illustrate there can be complicating factors. Linewidth estimates can be affected by instrumental noise that can decrease the accuracy of the fit. 

Our validation test confirmed that this linewidth criterion correctly identified the vast majority of airglow lines as ``extra-instrumental'', and the majority of cosmic ray spikes as '`intra-instrumental'', but it should not be considered definitive.  We use it as a `positive' criterion to identify candidates which should be given higher priority, but not a `negative' criterion that definitively rules them out.  The vast majority of our candidates that fail this final test are probably stealthy cosmic rays, but there may still be a small handful which are `extra-instrumental' emissions distorted by instrumental noise. The Tier 2 candidates that failed Tier 3 analysis (because their linewidth was below the predicted value for their given wavelength) are listed in Appendix \ref{s:rejects}.

\section{Results of Tier 1-3 Search}
\label{s:results}
After analyzing one spectrum each from our target list of stars, our search algorithm (Section \ref{s:automated}) identified 285 Tier 1 candidates which met our initial criteria and also did not match prominent terrestrial airglow lines. In Tier 2 analysis, 142 of these were manually identified as cosmic rays (Section \ref{s:cosmic}), 89 were in clusters indicative of natural stellar emission lines,  10 were identified as calibration lamp bleed-through (Section \ref{s:bleedthrough}), and 2 were too faint and noisy to properly categorize. This left us with 41 Tier 2 candidates. Of these, the curve-fitting algorithm (Section \ref{s:curvefit}) found 8 candidates with linewidths at or above the predicted instrumental FWHM and which therefore stood out as our highest priority Tier 3 candidates. 

\begin{table}
\begin{center}
\begin{tabular}{|l|l|} 
 \hline
 & Number of \\
 Category & Candidates \\ 
 \hline
 Cosmic Rays & 142 \\ 
 Natural Emission Line Clusters & 89 \\ 
 Calibration Lamp Bleedthrough &  10 \\ 
 Linewidth $<$ PSF & 34 \\
 Too dim to categorize & 2 \\
 Unexplained (Tier 3 candidates) & 8 \\
 \hline
 \end{tabular}
\end{center}
\caption{Classification of 285 candidates identified by the Tier 1 automated search algorithm, analyzing one observation per star.  This summarizes the results of the false-positive identification (Tier 2) and linewidth analysis (Tier 3) steps.}
\label{table:classification}
\end{table}

\subsection{Searching for Repeat Occurrences}
\label{s:repeat}
While laser transmissions could be ephemeral, signals that recur consistently would arguably be stronger candidates since they provide reproducible evidence and are less likely to be random noise.  Therefore, we repeated our analysis for {\em all} available observations of each of the 6 stars which yielded at least one Tier 3 candidate signal. Of 1,835 additional observations of these 6 stars, our initial search algorithm and curve fitting algorithm yielded 71 additional candidates at or above the predicted FWHM. We repeated our Tier 2 and 3 analysis from Sections \ref{s:falsepositive} and \ref{s:curvefit} on these observations, and categorized 29 as cosmic rays, 18 as clusters indicative of natural stellar emission lines, and 13 as due to calibration lamp bleedthrough. This left a total of 9 Tier 3 candidates for further analysis, 8 of which were in the original analysis and 1 new candidate which was found in the additional observations. In the case of the star that yielded a second candidate (HIP87607), the new candidate appeared at a different wavelength than the original candidate (see Table \ref{table:tier_3_candidates}).

\begin{table}
\begin{center}
\begin{tabular}{|l|l|} 
 \hline
 & Number of \\
 Category & Candidates \\ 
 \hline
 Cosmic Rays & 29 \\ 
 Natural Emission Line Clusters & 18 \\ 
 Calibration Lamp Bleedthrough &  13 \\ 
 Unexplained (in subsequent observations) & 1 \\
 Unexplained (in Total) & 9 \\
 \hline
 \end{tabular}
\end{center}

\caption{Categorization of additional spikes identified in additional observations of Tier 3 candidates from Table \ref{table:classification}.}
\label{table:followup}
\end{table}

\subsection{Assessment of Tier 3 Candidates}
\label{s:unexplained}
In total, 9 candidates passed the Tier 1-3 tests, three of which originated from the same star (HIP87607). In the case of the HIP87607 candidates, none repeated at the same wavelength. We later determined that the CD-312415 candidate does appear to have multiple emissions at nearby wavelengths, though none of the subsequent detections trigerred the automated search algorithm. The rest of our candidates did not repeat at all. However, regardless of their transient and non-repeating nature, they survived a battery of tests which eliminated many natural and anthropogenic false positives. Therefore, each one warranted further investigation. These Tier 3 candidates are listed in Table \ref{table:tier_3_candidates}.

\begin{table}
\begin{center}
\begin{tabular}{|c|c|c|c|} 
 \hline
& Spectral & & Wavelength\\
 Star & Type & Observation Time & \AA \\ 
\hline
 HIP59341 & G5V & 2013-06-01 03:04:25.906 & 4042.84\\
 CD-312415 & RG & 2018-02-11 01:10:40.597 & 3931.24\\ 
 GJ317 & M3.5V & 2018-04-10 03:25:08.275 &  4662.77\\ 
 HD127423 & G0V & 2013-06-01 04:16:19.375 &  5732.93\\
 HD96673 & K3V & 2006-01-28 07:17:55.086 & 3931.31\\
 HIP87607  & M0V & 2012-10-22 23:39:09.238 & 3784.09\\
 HIP87607  & M0V & 2012-10-22 23:39:09.238 & 4042.74\\
 HIP87607  & M0V & 2013-04-30 09:50:45.212 & 5454.02\\
 GJ4291 & K3V & 2005-07-20 04:08:04.468 & 3841.74\\
 \hline
 \end{tabular}
\end{center}
\caption{Final list of Tier 3 signals.}
\label{table:tier_3_candidates}
\end{table}

\section{Further Analysis of Tier 3 Candidates}

Having reduced the dataset to a handful of promising candidates, we subjected them to detailed scrutiny to identify sources of possible terrestrial contamination.

\subsection{Tier 4: Airglow Reanalysis}
\label{s:airglow_reanalysis}

Although the brightest airglow lines were automatically filtered out by our search algorithm, we re-analyzed the Tier 3 candidates in Table \ref{table:tier_3_candidates} to see if they matched any airglow lines in the more comprehensive database of \cite{faint_airglow_database}.  Care is required regarding Doppler shifts and reference frames. The HARPS data processing pipeline shifts the reduced HARPS spectra into the barycentric reference frame: we reverse that shift using the \texttt{HIERARCH ESO DRS BERV} parameter stored in the FITS header before comparing against the airglow database, to ensure we are comparing wavelengths in the terrestrial reference frame.

\begin{figure*}
    \centering  \includegraphics[width=\textwidth]{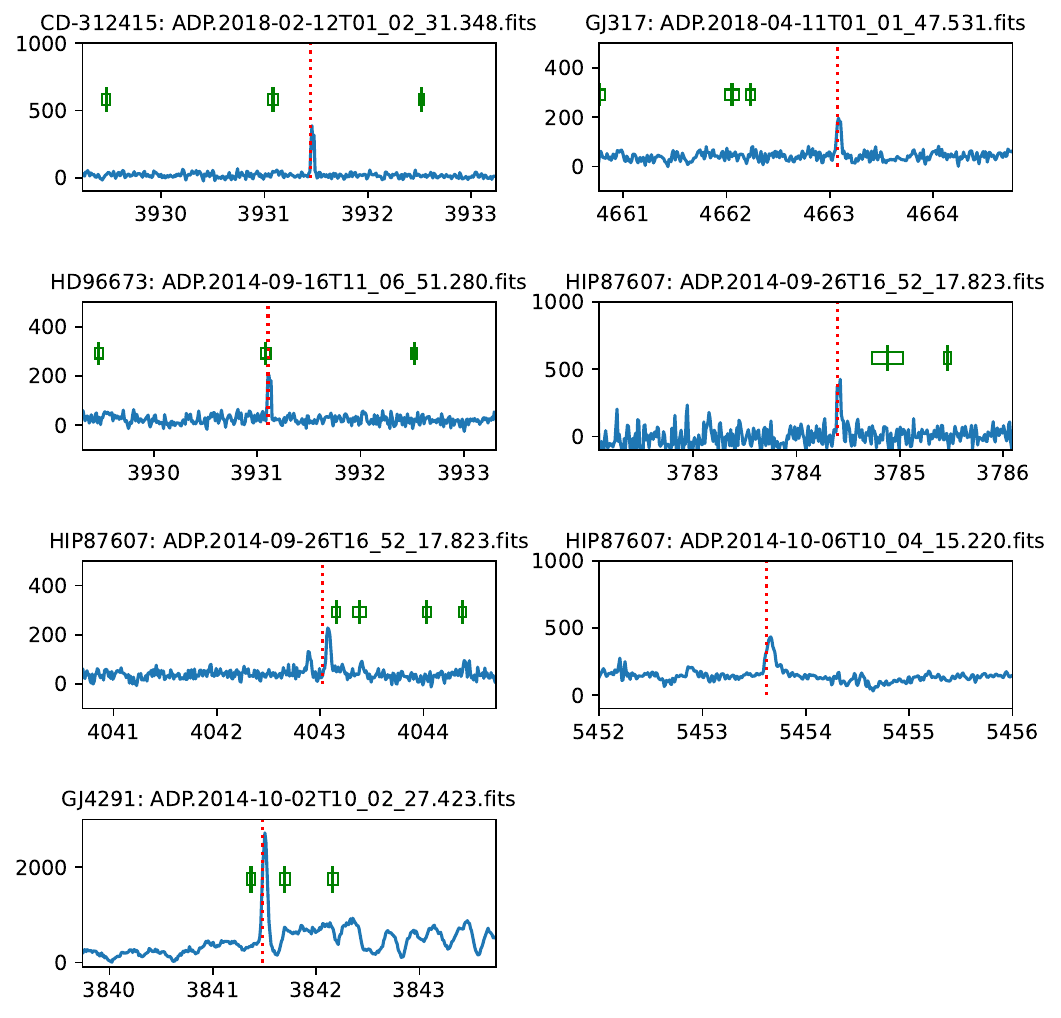}
    \caption{Comparison of Tier 3 candidate spikes from Table \ref{table:tier_3_candidates} against a detailed airglow line database.  Blue lines: reduced HARPS spectra, doppler-shifted into terrestrial reference frame.  Red dotted lines: candidate wavelength identified by our spike identification algorithm. Green boxes: central wavelength (line) and full width half maximum (box) for airglow lines listed in  \cite{faint_airglow_database}.}
    \label{fig:airglow_reanalysis}
\end{figure*}

Results are shown in Figure \ref{fig:airglow_reanalysis}. The HD96673 candidate is quite close to a known airglow line; the others definitely cannot be explained as airglow lines.

Sign errors are always possible with Doppler calculations.  The analysis above uses the same shifting method as was used to create Figure \ref{fig:stellarframespectra} below.  We confirmed that flipping the sign of the barycentric earth radial velocity did not give better concordance between candidate spikes and airglow lines. We also see in Table \ref{table:airglow_difference} that the barycentric correction velocity does not match the predicted velocity that would be necessary to account for the difference between the observed wavelength and the nearest airglow wavelengths, further arguing against airglow lines as an explanation.

\begin{table*}
\begin{center}
    \begin{tabular}{|c|c|c|c|c|c|c|}
    \hline 
    Star & Observation Date/Time  & $v_{\mbox{bary}}$ & $\lambda_{\mbox{bary}}$ & $\lambda_{\mbox{obs}}$ & $\lambda_{\mbox{airglow}}$ & $\Delta v$ (obs - airglow) \\
    & & (km/s) & (\AA) & (\AA) & (\AA) & (km/s) \\
    \hline
    CD-312415 & 2018-02-11, 01:10:40.597 & -15.983 & 3931.24 & 3931.03 & 
    3931.0825 & -3.96 km/s \\
    GJ317 &  2018-04-10, 03:25:08.275 &  -17.533 & 4662.77 & 4662.49 & 
    4662.0542 & 28.52 km/s \\
    GJ317 &  2018-04-10, 03:25:08.275 &  -17.533 & 4662.77 & 4662.49 & 
    4662.2324 & 17.06 km/s \\
    HIP87607 & 2012-10-22, 23:39:09.238 & -23.949 & 3784.09 & 3783.78 &
    3784.8796 & -86.54 km/s \\
    GJ4291 & 2005-07-20, 04:08:04.468 & 20.089 &  3841.74 & 3841.997 &
    3841.3704 & 48.95 km/s \\
    GJ4291 & 2005-07-20, 04:08:04.468 &  20.089 &  3841.74 & 3841.997 &
    3841.6980  & 23.37 km/s \\
    GJ4291 & 2005-07-20, 04:08:04.468 &  20.089 &  3841.74 & 3841.997 &
    3842.1606  & -12.75 km/s \\
    \hline
    \end{tabular}
\end{center}    

\caption{Comparison of candidate spectral signatures with nearby terrestrial airglow lines.  $v_{\mbox{bary}}$:  Relative velocity between solar barycenter and observatory, in direction of star.  $\lambda_{\mbox{bary}}$: wavelength of candidate spikes, in barycentric frame, as reported by HARPS spectral files.  $\lambda_{\mbox{obs}}$: wavelength of candidate spikes in observatory frame, calculated using  $v_{\mbox{bary}}$. 
 $\lambda_{\mbox{airglow}}$: wavelength of nearby airglow line. $\Delta v$: Velocity of airglow source relative to observatory that would be needed to explain candidate as an airglow line.  The large values for $\Delta v$ and inconsistency with the barycentric correction velocity indicate that these candidates cannot be explained by airglow.}
\label{table:airglow_difference}
\end{table*}

\subsection{Detailed Inspection of the HD127423 Candidate} 
\label{s:hd127423}

\begin{figure*}
    \centering  \includegraphics[width=\textwidth]{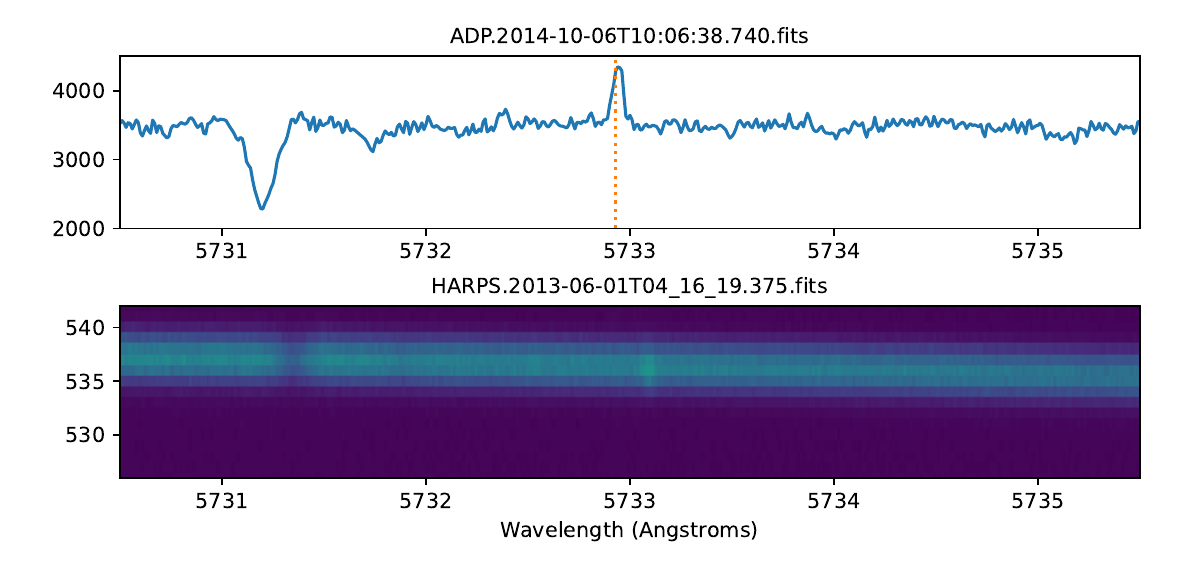}
    \caption{Star: HD127423, Spectral Type:G0V, Observation Date/Time: 2013-06-01, 04:16:19.375. Top panel: reduced spectrum.  Bottom panel: subsection of the corresponding CCD image for this wavelength range.}
    \label{fig:seti_candidate_HD127423}
\end{figure*}

The HD127423 candidate seemed especially interesting.  It appeared to originate from a G-Type star, had no nearby airglow lines, had a gaussian lineshape close to the instrument's FWHM, and, as seen in Figure \ref{fig:seti_candidate_HD127423}, the spectral energy is entirely within the CCD image of the spectrometer's entry slit, suggesting that this is not a cosmic ray or calibration lamp bleed-through.  We subjected it to additional detailed analysis.

The calibration lamp was on during the observation, and bleedthrough is observed at some wavelengths, but not at the candidate wavelength of 5732.93 \AA.

This candidate does not meet our criteria for natural stellar emission lines (Section \ref{s:stellaremission}).  The line occurs in isolation, and its linewidth is exceptionally narrow (just above the instrument's PSF). It is also spectral type G0V, which do not generally have emission spectra besides H and K lines \citep{HandKemissionlines}, which do not match the candidate's wavelength. 

The candidate does not appear in our list of major airglow lines (Table \ref{table:AirglowWavelengths}), nor does it match any of the fainter airglow lines listed in the UVES database \citep{faint_airglow_database} \citep{UVES_database}. 

We considered the possibility of human laser interference by overflying LiDAR satellites and drones.  A review of LiDAR technologies found that our candidate at 573.3 nm does not match the wavelengths used by LiDAR satellites (532 nm, 355 nm, and 1064 \citep{lidar_history}), airborne LiDAR systems (1064 nm, (op. cit.)), autonomous vehicles (905 nm and 1550 nm \citep{vehicle_lidar}), and oceanic bathymetric LiDAR systems (532 nm, \citep{bathymetric_lidar}). 

We also considered interference from ground-based adaptive optics lasers. However, ESO's adaptive optics lasers excite atmospheric sodium lines at 589 nm, which does not match our candidate and which would have been filtered out as an airglow line \citep{HARPS_guidestars}. In general, Laser Guidestar Adaptive Optics (LGAO) systems work by artificially inducing airglow \citep{LaserGuideStarAdaptiveOptics}, and would be ignored by our airglow rejection filter.  Other adaptive optics systems can utilize Raleigh scattering instead of artificially induced airglow--but these systems generally operate at blue wavelengths \citep{RaleighScattering} rather than the yellow/green range of the HD127423 candidate.

Neither the HD127423 candidate, nor any of the other final candidates, fit these categories of anthropogenic false positives (LiDAR satellite/drone interference and adaptive optics lasers, etc.)

The HD127423 candidate is not a frequently observed feature that might represent an instrumental artifact, spurious byproduct of the blaze function or reduction pipeline error. Analysis by Nikolai Piskunov (pers. comm.) noted that no recurring spike at this wavelength was observed in a random sample of 396 HARPS spectra. 

However, there is also overwhelming evidence that this signature could not have arrived from interstellar distances.  

The spike in question only appears once in all 8 observations of HD127423 available in the HARPS database.  However, as noted by N. Piskunov (pers. comm.), it appears in observations of other stars taken on the same night, including HD107181, HD128356, HD126535, HD143120, HD143361, HD144899.  Each of these observations showed a spike at or near 5733\AA, even though the stars were separated by hours of right ascension. The eight observations were taken within a two-hour window, with a 15-minute exposure time for each consecutive observation. The odds that extraterrestrials beaming lasers from stars at such varied distances would all happen to have their signals arriving at Earth at the same time seemed small. 

\begin{figure*}
    \centering
    \includegraphics[width=\textwidth]{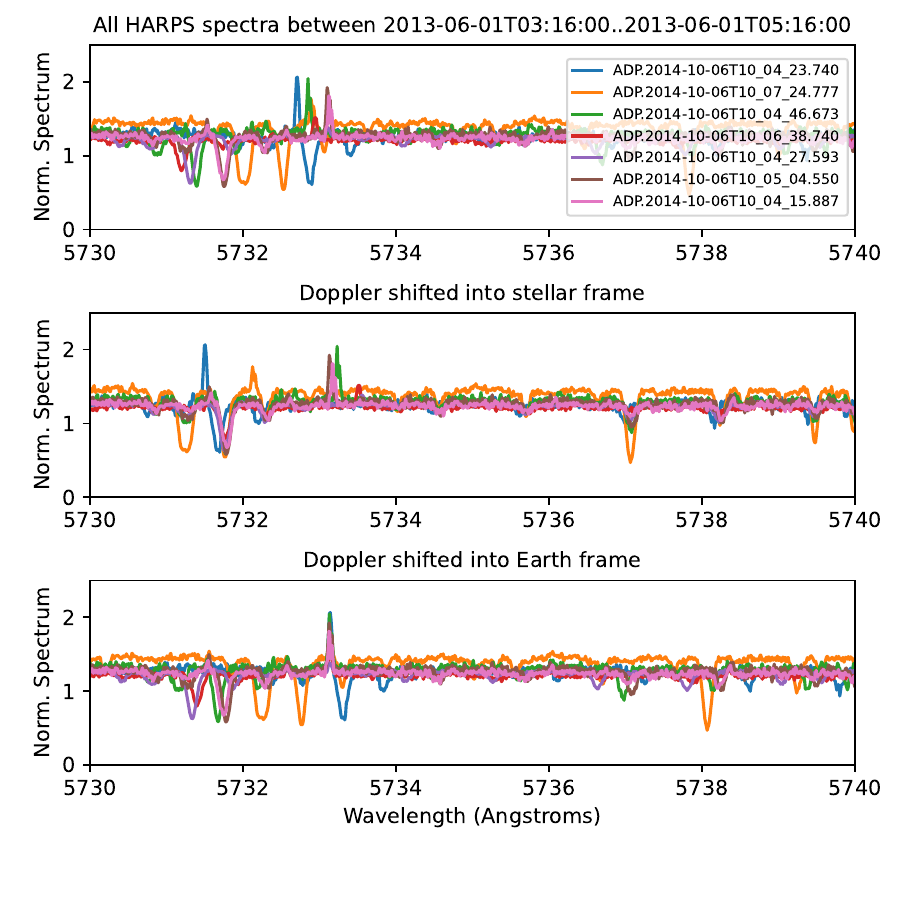}
    \caption{Reduced spectra for all stars observed by HARPS within +/- 1 hour of the HD127423 candidate at 2013-06-01, 04:16:19.375 (HD127423, HD107181, HD128356, HD126535, HD143120, HD143361 and HD144899).  Each observation showsa spike around 5733 \AA. Top panel: all 7 spectra in the default reference frame for HARPS pipeline (solar system barycenter). Middle panel: all 7 spectra, doppler shifted to the reference frame of each respective star. Bottom panel: all 7 spectra, Doppler shifted into the Earth's reference frame.  The alignment of the 5733{\AA} spikes in the terrestrial reference frame indicates that this is a terrestrial signal, not a stellar feature.}
    \label{fig:stellarframespectra}
\end{figure*}

To confirm that this was a terrestrial signal, we Doppler-shifted each spectrum into its respective stellar reference frame (Figure \ref{fig:stellarframespectra}).  We observed that absorption features common to all these stars align, but the spike does not. Thus, the signal is not in the stars' reference frames.  Second, we Doppler-shifted each spectrum into the {\em terrestrial} observatory's reference frame using the barycentric radial velocity listed in the using the \texttt{HIERARCH ESO DRS BERV} FITS header, and confirmed that the stars' absorption features do not line up, but the mysterious spike does.

 We confirmed that the spike was in the observatory's reference frame by looking at the raw CCD images, which showed a common spike at the same CCD pixels in all of them.  There was no evidence of CCD "bleed-through" at this location, and these features appear at the same pixel position on the CCD images, confirming that there is no doppler shift at all relative to the calibration source. Thus, our analysis suggests that this is an as yet unknown terrestrial source.

The two-hour time window during which the source persisted argues against satellites in low earth orbit, since at such velocities the satellite would not stay in the same location long enough to continuously beam down a laser for that duration. However, we can also rule out a transmitter at interstellar distances. We were therefore left with a signature which is unexplained, yet originating from a local source. 

\subsection{Tier 5: Doppler Detection of Unknown Terrestrial Sources}
\label{s:doppler}
Having identified an unknown terrestrial source for HD127423 (Section \ref{s:hd127423}) by looking for the same spike in contemporaneous observations,  we repeated this process for the other Tier 3 candidates.  After examining observations taken within a two hour window on the same night as each of our candidates, we found 3 more candidates in which also show up in other same-night observations and appear to be stationary in the Earth's reference frame: these are therefore also terrestrial in origin.  Although no longer viable SETI candidates, since they do not match known airglow lines they may represent novel atmospheric phenomena potentially of interest to meteorologists. Thus, we are left with four observations which are unknown terrestrial sources, and three Tier 5 candidates which may represent unknown extraterrestrial sources (see \ref{tab:terrestrial_or_extraterrestrial}). Although the HD96673 candidate is conspicuously near the wavelength of a faint airglow line (see \ref{s:airglow_reanalysis}), it only appeared in one of the ten observations taken that night.  

\begin{table}
    \centering
    \begin{tabular}{|c|c|c|}
    \hline
        Observation Date/Time  & Star & Terrestrial Source? \\
    \hline
       2013-06-01T03:04:25.906 & HIP59341 & YES \\
       2013-06-01T04:16:19.375 & HD127423 & YES \\
       2013-04-30T09:50:45.212 & HIP87607 & YES \\
       2005-07-20T04:08:04.468 & GJ4291 & YES \\
       2018-04-10T03:25:08.275 & GJ317 & NO (Tier 5)\\
       2012-10-22T23:39:09.238 & HIP87607 & NO (Tier 5)\\
       2018-02-11T01:10:40.597 & CD-312415 & NO (Tier 5)\\
    \hline
    \end{tabular}
    \caption{In Tier 5 analysis, each Tier 4 candidate was tested to see whether similar emission features appears in observations of other stars on the same night, and if so whether these emissions are stationary in the terrrestrial frame.  Candidates that do not appear to have a persistent terrestrial source form our final Tier 5 candidate list.}
    \label{tab:terrestrial_or_extraterrestrial}
\end{table}

Over half of our Tier 4 candidates originate from M-dwarf stars. These lines occurred in isolation, rather than as one of a ``forest'' of broader natural lines. However, as discussed in Section \ref{s:stellaremission}, these stars can sometimes produce relatively narrow emission lines due to relatively low temperatures, high metallicity, and frequent flaring activity \citep{Marcy_2021}. Therefore, though they are unexplained, a natural origin cannot be definitively ruled out. Conversely, the candidates from G and K type stars initially stood out as the most compelling, since such stars are generally not known to produce many natural emission lines \citep{stellar_emission_linesJ}. However, the Tier 5 analysis revealed that all of our G and K-type candidates appear to be from unknown terrestrial sources which are stationary in the telescope's reference frame..

\begin{figure*}
    \centering  \includegraphics[width=\textwidth]{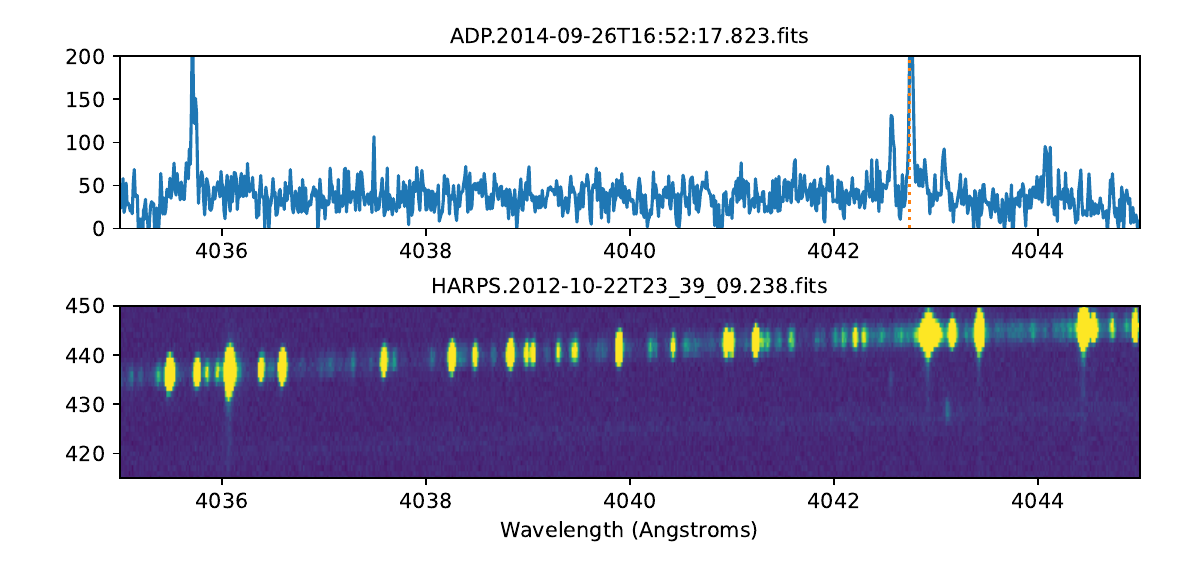}
    \caption{Star: HIP87607, Spectral Type: M0V, (Observation Date/Time: 2012-10-22, 23:39:09.238) Top panel: reduced spectrum. Bottom panel: subsection of the corresponding CCD image for this wavelength range.}
    \label{fig:seti_candidate_HIP87607}
\end{figure*}

\begin{figure*}
    \centering  \includegraphics[width=\textwidth]{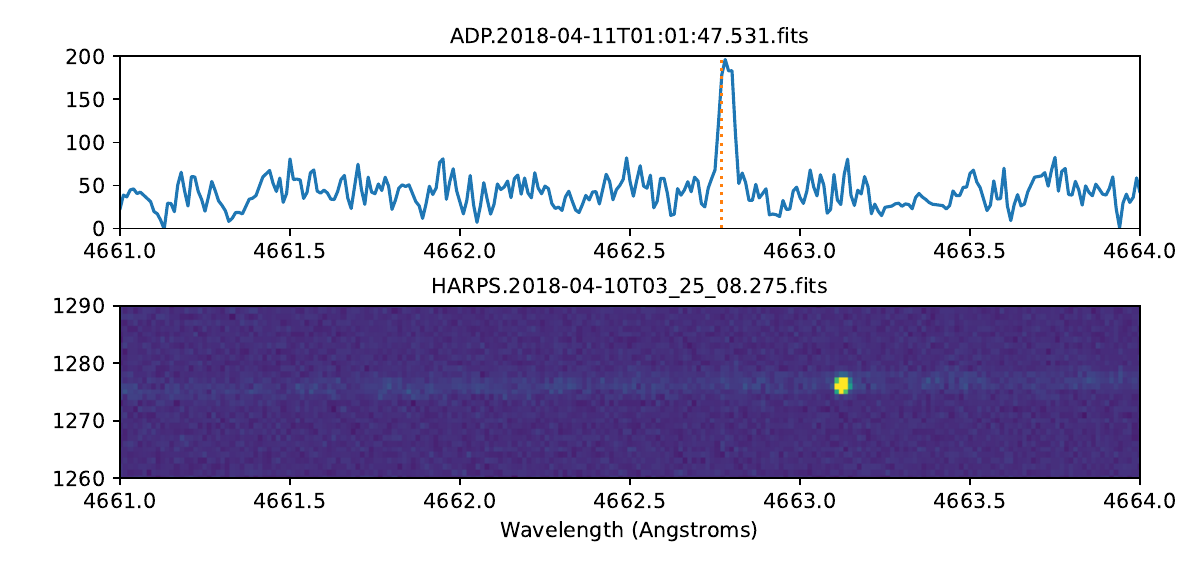}
    \caption{Star: GJ317, Spectral Type: M3.5V, (Observation Date/Time: 2018-04-10, 03:25:08.275) Top panel: reduced spectrum. Bottom panel: subsection of the corresponding CCD image for this wavelength range.}
    \label{fig:seti_candidate_GJ317}
\end{figure*}

A total of three Tier 5 candidates currently show no indication of being from a terrestrial source and do not match our criteria for any of our false positive categories, therefore warranting further investigation. These include a candidates from M-type stars HIP87607 (see: \ref{fig:seti_candidate_HIP87607}) and GJ317 (see: \ref{fig:seti_candidate_GJ317}), as well as from the oscillating red giant CD-312415.

\section{Estimating the Power Requirements for a Transmitting Laser}
If one of our candidate signatures came from an extraterrestrial laser, how powerful would it need to be for us to detect it?  We will use the GJ317 spike as a benchmark example.  Our calculation follows \cite{Lipman_2019}.  We do not know the diameter of the aperture of the laser's transmitting optics: in these calculations, we assume 10 meters, similar to our own largest telescopes.

The laser beam will spread out by diffraction as it travels to Earth, with an opening angle:

\begin{equation}
    \theta = (\frac{1.22\lambda}{2d_T})
\end{equation}
where $\lambda$ is the wavelength of the beam and $d_T$ the diameter of the transmitter's aperture.  On reaching Earth, the beam will be spread out into a disk of area

\begin{equation}
    A = \pi r^2 = \pi (\frac{\theta}{2} D)^2 = \pi \frac{1.22^2 \lambda^2 D^2}{4 d_T^2}
\end{equation}
where $D$ is the distance from the transmitter to Earth.  If the transmitting laser's luminosity is $L_T$, the luminous flux at Earth is

\begin{equation}
    F = \frac{L_T}{A} = \frac{4 L_T d_T^2}{\pi 1.22^2 \lambda^2 D^2}
\end{equation}
The power, in watts, entering the receiving telescope is thus

\begin{equation}
    P_R = A_R F = \pi (d_R/2)^2 F = \frac{L_T d_T^2 d_R^2}{1.22^2 \lambda^2 D^2}
\end{equation}
where $A_R$ is the area and $d_R$ the diameter of the receiving telescope.  The number of photons entering the telescope over the course of an observation of duration $T$ is thus

\begin{equation}
    N = P_R T \frac{\lambda}{h c} = \frac{L_T d_T^2 d_R^2 T}{1.22^2 \lambda h c D^2}
\end{equation}
where $h$ and $c$ are Planck's constant and the speed of light.  Accounting for the instrument's quantum efficiency $\epsilon$, the number detected will be
\begin{equation}
    N_{\mbox{det}} = \frac{\epsilon L_T d_T^2 d_R^2 T}{1.22^2 \lambda h c D^2}
\end{equation}
Solving for the transmitter power $L_T$:
\begin{equation}
    L_T = \frac{1.22^2 N_{\mbox{det}} \lambda h c D^2}{\epsilon d_T^2 d_R^2 T}
\end{equation}

\begin{table}[h]
\begin{tabular}{ll}
$N_{\mbox{det}}$ & 1500\\
$\lambda$ & 4663 \AA \\
D & 15.2 pc \\
$\epsilon$ & 0.057 (from HARPS user manual)\\
$d_T$ & 10m (assumed) \\
$d_R$ & 3.6m \\
$T$ & 900 s\\
\end{tabular}
\label{t:powervalues}
\caption{Values used to calculate transmitter power needed to explain spectral spike candidate in observation of GJ317 at 2018-04-11, 01:01:47.531.}
\end{table}

Plugging in values given in Table \ref{t:powervalues} using the GJ317 candidate as an example, we get a required transmitter power of about 700 W, continuously for the duration of the 15 minute exposure time. If the transmitter were pulsed, higher power would be needed, in proportion to the transmitter's duty cycle. 

This calculation does not factor in extinction due to interstellar dust.  \cite{Hippke_2018} estimated power losses due to astronomical dust, and found them to be appreciable only for distances $>=$ 300 pc, and an order-of-magnitude problem at distances $>=$ 1000 pc.  Thus, we are justified in ignoring them for close targets like GJ317 (15.2 pc).

Thus, using the GJ317 candidate as a benchmark, we find that this survey has a sensitivity capable of detecting lasers which are possible to manufacture with currently existing human technology \citep{Lander1997ContinuouswaveCD}.

\section{Summary and Conclusions}
We analyzed single archived reduced spectra from 2821 stellar targets taken by the ESO’s HARPS spectroscope to search for potential signatures of lasers from extraterrestrial civilizations. We identified false positives (Section \ref{s:falsepositive}), including cosmic rays (Section \ref{s:cosmic}), bleed through from the Th-Ar calibration lamp (Section \ref{s:bleedthrough}), night sky airglow lines (Section \ref{s:Airglow}), and large clusters of emission lines which are likely the product of natural stellar activity (Section \ref{s:stellaremission}). Of the 2821 stellar spectra we analyzed, we identified 285 Tier 1 candidates with at least one narrow spectral emission feature.  41 of these did not fit any of the above false positive categories, becoming Tier 2 candidates. We filtered this dataset further on the basis of linewidth (Section \ref{s:curvefit}) and analyzed all other HARPS observations of these stars (Section \ref{s:repeat}). We were left with a total of 9 compelling Tier 3 candidates (Section \ref{s:unexplained}). After further investigation, we identified one that could be a faint airglow line (Section \ref{s:airglow_reanalysis}, Tier 4). We then analyzed observations from other stars taken on the same night as each candidate within a two hour window. After shifting these additional spectra into the terrestrial reference frame, we found that four of the surviving candidates originate from unknown sources which are stationary in the telescope's reference frame (See 
\ref{tab:terrestrial_or_extraterrestrial} in \ref{s:hd127423}). Although no longer viable SETI candidates and definitely terrestrial in origin, the source remains unknown. The two hour window during which they appeared argues against a laser from a classified satellite in Low Earth Orbit, though classified satellites in Geostationary cannot yet be ruled out. They also might represent novel atmospheric phenomena and could be of interest to meteorologists. This leaves three Tier 5 candidates which passed through all of our filters for prosaic explanations and show no indication of being terrestrial in origin, and therefore warrant further investigation as SETI candidates.

Of our surviving Tier 5 candidates, two of them (GJ317 and HIP87607) originate from M-type stars, and one of them (CD-312415) originates from an oscillating red giant \citep{CD-312415_RED_GIANT}. Although these three candidates do not match our criteria for stellar emission lines, detection of candidate laser signals is more ambiguous in red dwarfs due to their relatively cool temperatures, high metallicity, and coronal activity, which leads to flaring and prominent narrow emission lines \citep{Marcy_2021}. Similarly, an oscillating red giant star also indicates stellar activity at relatively cool temperatures, which makes it a much more ambiguous candidate. Nonetheless, they still warrant further followup. 

Since our study analyzed only one observation from each target and then performed follow-up analysis of additional observations only on stars where a candidate was found, it can only constrain the likelihood of continuous sources repeatedly directed toward Earth.  It could easily miss more intermittent signals. Thus, our study must contend with the longstanding challenge faced by many other SETI searches: how can we adapt our methodologies to account for the possibility of intermittent transmitters or ``one-off'' signals?  Can ``one-off'' signals constitute SETI detections, or is repeatability an inescapable requirement of a confirmed detection \citep{2019nasatechnosignatures, Wow!Signal}? These questions will define the future direction of this project.

\section{Future Avenues of Research}
\subsection{Expanding the Search}
Our search only analyzed one observation from each star. While we did look at all subsequent observations for the handful of stars which yielded compelling candidates, time constraints limited us to survey only a small fraction of the total number of observations.  In future work, we will extend the search to analyze every stellar observation in the HARPS archive, which will reveal more candidate signals and may allow us to better investigate the possibility of intermittent signals. To make this task feasible, we will incorporate the Doppler-shifted airglow identification in Section \ref{s:airglow_reanalysis} and the curve-fitting analysis of Section \ref{s:curvefit} into the initial automated search. Ancillary data from the HARPS pipeline may also allow us to automate cosmic ray detection.

We can also expand our search spatially, by covering the Northern Hemisphere. HARPS-N operates as a sister instrument with the same capabilities \citep{HARPS_N}, but covering a different set of stars. Therefore, we can expand the search both spatially and temporally to include orders of magnitude more observations and targets, improving our odds of discovering technosignatures and novel astrophysical anomalies.

\subsection{Further Investigation of the Tier 5 Candidates}

We will also continue to investigate the final three Tier 5 candidates. In particular, since these candidates are from M-Type and oscillating Red Giant stars, the key challenge will be ascertaining whether they represent natural stellar activity. We will attempt to develop rigorous tests to differentiate natural astrophysical stellar events from potential technosignatures. 

Additionally, initial manual analysis of CCD images and subsequent curve fitting algorithm is sufficient to rule out a vast majority of cosmic rays. However, there still may be some rare scenarios which allow cosmic rays to slice through the CCD at an angle which leaves little to no visible signature in the final image and happens to match or exceed HARPS' PSF/FWHM. We intend to use optimal extraction \citep{horne86optimal_extraction} as an additional test to determine if the pattern of these features extracted directly from the eschellogram  match the expected pattern from starlight passing through the HARPS optics. The ancillary data from the HARPS pipeline \citep{HARPS_specs} may provide optimal extraction data: if not, a bespoke reanalysis will be performed. 

We will also investigate whether any of these three candidate stellar systems stand out as unusual in other datasets in ways which may further substantiate the ETI origin hypothesis, such as the presence of potentially habitable exoplanets \citep{exoplanets}, and perhaps even biosignature / biosignature gasses in the atmospheres of those planets \citep{exoplanetbiosignatures,HAQQMISRA2022194}, detection of unusual radio emissions \citep{BACKUS1998651}, infrared excess that might indicate system-scale megastructures \citep{dysonspheres,TILGNER1998607}, or indications of nuclear waste dumped in the stellar atmospheres \citep{WHITMIRE1980149}. In the case of our oscillating Red Giant candidate (CD-312415), habitability is  unlikely. Still, analysis of other anomalies can either point to other corroborating technosignatures or (more probably) natural stellar characteristics which may be responsible for the emission. In the case of the GJ317 and HIP87607 candidates, we need to develop more refined tools to distinguish M-type flaring activity from potential lasers. Finally, we hope that our presentation of these candidates here inspires other investigators to take a closer look at them as well.

By expanding our search temporally and spatially, improving our automated detection methods, considering whether our three surviving candidates are natural stellar activity or potential technosignatures and determining if they stand out in other types of complementary SETI searches, and inviting others to perform follow-up analysis, we intend to continue to assess the rich optical SETI dataset provided by HARPS. Thus, this paper does not represent an endpoint-rather, we hope it is only the start of a much larger inquiry,

\section{Acknowledgements}
This project is based on data products from observations made with ESO Telescopes at the La Silla Paranal Observatory. We would like to thank Nikolai Pishkunov and Jason T. Wright for their helpful suggestions and input in our analysis of the HD127423 candidate. We also thank Dan Wertheimer for giving us the idea to do this project in the first place. Additionally, this project began as an undergraduate thesis--so we thank the rest of the thesis committee: Dipankar Maitra, Michael Drout, and Geoffrey Collins. Finally, we thank Lew Levy and the SETI Institute's Forward Award committee for conference travel support.

Both authors declare that they have no conflicts of interest.

\appendix

\section{Tier 2 Candidates Rejected by the Curve Fitting Algorithm (Tier 3)}
\label{s:rejects}

\begin{table}[h!]
\begin{center}
\begin{tabular}{|c|c|c|c|} 
 \hline
 Star & Spectral Type & Observation Time & Wavelength \\ 
 \hline
 BD-194341 & K0IV & 2015-07-20, 01:34:29.692 & 6693.13 \AA \\
 CoRoT-22 & G8V & 2010-06-07, 03:31:53.279 & 6864.20 \AA \\
 GJ3200 & G3V & 2015-09-17, 09:25:58.011 & 6717.03 \AA \\
 GJ3436 & G7VFe-1.5 & 2012-11-27, 07:50:10.982 & 6136.16 \AA \\
 GJ493 & K5.5V & 2013-02-04, 08:24:36.477 & 6473.91 \AA \\
 GJ722 & G6V & 2006-03-14, 07:34:56.654 & 6011.96 \AA \\
 GJ9075 & F8V & 2015-09-17, 08:11:02.767 & 6294.7 \AA \\
 GJ9322 & G3/5V & 2006-04-09, 02:04:21.684 & 6850.47 \AA \\
 HD104800 & G3V & 2015-01-14, 08:15:07.195 & 6212.84 \AA \\
 HD128207 & B8V & 2015-01-21, 07:49:50.149 & 4252.94 \AA \\
 HD134088 & G2/3(V) & 2016-03-15, 06:11:36.659 & 6826.23 \AA \\
 HD138403 & O8(f)ep & 2006-03-24, 06:48:41.041 & 6355.8 \AA \\
 HD145689 & A3V & 2006-05-21, 03:03:28.159 & 6225.54 \AA \\
 HD147644 & F8V & 2015-04-19, 06:48:21.114 & 5937.49 \AA \\
 HD149933 & K0V & 2005-05-14, 05:50:20.610 & 5794.54 \AA \\
 HD157060 & F9V & 2015-04-19, 06:07:40.929 & 6650.82 \AA \\
 HD112109 & F0V & 2018-04-21, 04:01:22.672 & 4840.94 \AA \\
 HD184711 & G5wA/F & 2008-09-20, 00:41:44.834 & 6017.72 \AA \\
 HD186194 & G3V & 2013-06-01, 07:45:16.866 & 5733.43 \AA \\
 HD186194 & G3V & 2013-06-01, 07:45:16.866 & 5817.84 \AA \\
 HD205289 & F5VFe-0.8CH-0.4 & 2004-09-22, 01:25:01.520 & 6687.43 \AA \\
 HD39091 & G5V & 2019-04-27, 22:52:34.109 & 5925.82 \AA \\
 HD43940 & K3V & 2006-09-11, 07:41:55.332 & 5422.82 \AA \\
 HD47186 & G5V & 2016-01-13, 06:55:06.378 & 6650.75 \AA \\
 HD56413 & G0V & 2015-11-12, 07:07:02.375 & 6769.09 \AA \\
 HD68284 & G8/K0V & 2013-11-29, 08:30:47.035 & 6715.78 \AA \\
 HD71479 & ApEuSr & 2006-01-24, 04:28:30.077 & 6833.37 \AA \\
 HD90926 & G6V & 2007-02-10, 06:17:51.971 & 6274.12 \AA \\
 HD96423 & G5V & 2012-05-01, 02:26:12.958 & 6121.38 \AA \\
 HD98248 & G0 & 2005-05-16, 02:18:59.000 & 6763.27 \AA \\
 HD77191 & ApSrEu(Cr) & 2004-11-23, 08:38:23.489 & 6445.04 \AA \\
 TYC6296-1556-1 & N/A & 2014-06-24, 06:47:40.911 & 6712.83 \AA \\
 TYC6822-2535-1 & N/A & 2017-10-11, 00:13:56.459 & 6341.24 \AA \\
 TYC8233-68-1 & N/A & 2017-05-17, 04:15:56.151 & 5651.96 \AA \\
 \hline
 \end{tabular}
\end{center}
\caption{Tier 2 candidate spikes that failed Tier 3 analysis because their line widths were narrower than the the predicted FWHM of the instrument.  As discussed in Section \ref{s:curvefit}, this indicates that these candidates are probably not ``extra-instrumental'', but we include them here for completeness.}
\end{table}
\clearpage

\bibliographystyle{aasjournalv7}
\bibliography{references.bib}

\end{document}